\documentclass[twocolumn,reprint,superscriptaddress,amsmath,amssymb,aps,pra]{revtex4-2}
\usepackage{units}
\usepackage{graphicx,subfigure}
\usepackage{mathrsfs}
\usepackage{bm}% bold math
\usepackage{textcomp}
\usepackage{url}
\usepackage{dsfont} 
\usepackage{braket}
\usepackage[compact]{titlesec}
\usepackage[colorlinks=true,linkcolor=magenta,citecolor=blue]{hyperref}%
\usepackage{todonotes}
\usepackage{empheq}
\usepackage{multirow}
\begin{document}
\title{Goos-H\"{a}nchen Shift with a rotating atomic superfluid in a ring}
\author{Ghaisud Din}
\affiliation{Ministry of Education Key Laboratory for Nonequilibrium Synthesis and Modulation of Condensed Matter,
Shaanxi Province Key Laboratory of Quantum Information and Quantum Optoelectronic Devices, School of Physics, Xi’an Jiaotong University, Xi’an 710049, China}
\author{Muqaddar Abbas}
\email{muqaddarabbas@xjtu.edu.cn}
\affiliation{Ministry of Education Key Laboratory for Nonequilibrium Synthesis and Modulation of Condensed Matter,
Shaanxi Province Key Laboratory of Quantum Information and Quantum Optoelectronic Devices, School of Physics, Xi’an Jiaotong University, Xi’an 710049, China}
\author{Pei Zhang}
\email{zhangpei@mail.ustc.edu.cn}
\affiliation{Ministry of Education Key Laboratory for Nonequilibrium Synthesis and Modulation of Condensed Matter,
Shaanxi Province Key Laboratory of Quantum Information and Quantum Optoelectronic Devices, School of Physics, Xi’an Jiaotong University, Xi’an 710049, China}
%%%%%%
%%%%%%
\date{\today}
\begin{abstract}
We investigate the Goos--H\"{a}nchen shift of the transmitted probe and show that optomechanical interference in a ring Bose-Einstein condensate provides a sensitive, rotation–tunable beam–shift response at ultralow optical powers. Without quantized circulation, conventional optomechanically induced transparency produces a strictly positive and bounded shift whose magnitude is governed primarily by cooperativity; small detuning offsets can introduce a weak, transient sign change consistent with a Fano-type asymmetry, but the overall response remains limited. With circulation, the Bragg-scattered mechanical side modes split, yielding a double–transparency dispersion with steep dispersive flanks that strongly amplify the phase derivative and bias its sign. In this regime, the peak shift grows monotonically with control–field strength, reflecting enhanced linearized coupling and increased transmission across the angular scan. At fixed power, the detuning dependence is decisive: the shift is maximized at the red–sideband condition and diminishes away from resonance, tracking how effectively the scan samples the rotation–split dispersive flanks. Increasing the winding number broadens the central absorption and steepens accessible phase gradients, further boosting the attainable shift. The protocol remains minimally invasive under experimentally realistic conditions, and interatomic interactions have a negligible influence on the transmission features that set the phase slope. These results identify circulation, control power, and cavity detuning as practical knobs for in situ control of the Goos--H\"{a}nchen shift, enabling interferometric beam steering and phase–gradient metrology in hybrid atom–optomechanical platforms.
\end{abstract}
\maketitle
\section{Introduction}
\label{Sec:intro}
 Over the past several years, Bose--Einstein condensates (BECs) confined in ring, have become a workhorse setting for probing a wide variety of effects~\cite{PhysRevLett.110.025302,PhysRevLett.111.205301,PhysRevA.91.013602,PhysRevLett.123.250402} that arise from quantized circulation~\cite{RevModPhys.81.647}. In a closed loop, the condensate order parameter must be single valued, forcing the phase to wind by integer multiples of $2\pi$; this constraint supports persistent currents that can survive for macroscopic times~\cite{freilich2010real,PhysRevLett.99.260401,cooper2008rapidly}. Certain configurations even host topologically robust states with larger circulation numbers~\cite{PhysRevLett.93.160406}. Because of these features, ring-trapped BECs serve as an adaptable platform for studying superfluid hydrodynamics~\cite{tilley2019superfluidity,mehdi2021superflow}, implementing matter-wave interferometers~\cite{PhysRevLett.126.170402}, building elementary atomtronic circuit elements~\cite{amico2021roadmap,PhysRevLett.121.090404}, and investigating aspects of fermionic superfluidity~\cite{PhysRevX.12.041037,PhysRevLett.128.150401,PhysRevLett.121.225301}.

Across these applications, accurately determining how much circulation is present in essential. Operationally, the rotation of the atomic superfluid is encoded in the phase winding number $m$ of the macroscopic wavefunction; thus, measuring rotation amounts to extracting this integer $m$. Most commonly used techniques infer the winding number via absorption imaging of the atoms, a procedure that is intrinsically destructive: the measurement removes or strongly perturbs the sample, hindering real-time monitoring and forcing careful reproduction of the same initial conditions between shots~\cite{PhysRevLett.110.025302}. A further practical obstacle is geometric. The relevant length scale---the radius of the vortex core or the narrow width of the ring is typically smaller than an optical wavelength, which makes the direct in situ visualization of vortices in dilute trapped gases difficult. Consequently, experiments often allow the cloud to expand before imaging so that the circulation signatures become resolvable, relying on time-of-flight techniques~\cite{eckel2014hysteresis}.

 To address the limitations discussed above, a recent proposal~\cite{PhysRevLett.127.113601} introduces a flexible architecture that can both sense and actively control the circulation of a ring-shaped Bose-Einstein condensate with only minimal disturbance to the sample. The scheme is designed for a truly in situ and real-time operation, allowing the circulation state to be continuously read while the atoms remain trapped~\cite{PhysRevLett.127.113601}. The core mechanism is cavity optomechanics~\cite{RevModPhys.86.1391}: atoms inside a high-finesse optical resonator couple dispersively to one or more cavity modes, so that collective motion (and in particular rotational degree of freedom) imprints a measurable modulation on intracavity light. In turn, the intra-cavity field exerts a radiation--pressure--like backaction on the atoms, providing a coherent pathway for manipulation and feedback control~\cite{RevModPhys.86.1391}. Compared with previously demonstrated techniques, this approach yields an improvement in the rotation sensitivity of ring BECs by approximately three orders of magnitude (a factor of $\sim 10^3$)~\cite{PhysRevLett.127.113601}. Beyond enhanced metrology, the same light-matter coupling provides a route to steer atomic currents by generating optomechanical entanglement between the optical field and the matter-wave modes of the condensate, allowing coherent control of circulation states~\cite{PhysRevLett.127.113601,RevModPhys.86.1391}.

On the other hand, the Goos-H\"{a}nchen (GH) shift\cite{PhysRevE.66.067603,PhysRevA.81.023821,PhysRevLett.111.223901} is the small longitudinal displacement experienced by a bounded light beam upon reflection from an interface: rather than returning from the geometrical point of incidence, the beam centroid emerges shifted along the plane of reflection~\cite{goos1947neuer,goos1949neumessung}. Within a wave-optics view, the effect follows from the frequency- and angle-dependent reflection phase; the displacement is proportional to the derivative of this phase with respect to the in-plane wave vector (the Artmann relation)~[3]. The GH shift appears under total internal reflection as well as at metal, dielectric, and hybrid interfaces, and it can be strongly modified by multilayer stacks, plasmonic/photonic resonances, metamaterials, and two-dimensional media~\cite{artmann1948berechnung,renard1964total,puri1986goos}. Depending on the underlying dispersion, both positive and negative shifts have been reported, with pronounced enhancements near sharp spectral features or critical coupling conditions~\cite{aiello2008role}. Beyond being a fundamental beam-shaping phenomenon, the GH shift has found utility in precision metrology, refractive-index and thickness sensing, and beam-steering elements; weak-value amplification and other interferometric readout schemes have further boosted its detectability~\cite{bliokh2013goos,PhysRevLett.68.931}.

 Cavity optomechanical platforms offer a particularly attractive route to engineer the GH effect. In such systems, radiation-pressure coupling between an optical mode and a mechanical (or collective) degree of freedom produces steep, controllable phase dispersion and interference features (e.g., transparency and absorption windows) that directly set the magnitude and sign of the GH displacement for a reflected or transmitted probe~\cite{PhysRevA.100.063833,PhysRevA.102.053718}. By tuning the pump–probe detuning, optical power, and linewidths, prior studies have demonstrated large, tunable, and even sign-reversed GH shifts in optomechanical resonators~\cite{PhysRevA.100.063833,PhysRevA.102.053718}.

 Despite these advances, the GH shift has not, to the best of our knowledge, been investigated in the current platform considered here—namely, a ring-trapped Bose--Einstein condensate coupled to an optical cavity that exhibits optomechanically induced transparency and related interference phenomena.

Previous investigations of ring-trapped Bose-Einstein condensates in cavity optomechanical system\cite{PhysRevA.107.013525} have already established central features relevant to this work. They, investigate under resonant driving, narrow, symmetric transparency windows in the probe transmission. In the limit of zero quantized circulation, the response reduces to the well-known phenomenon of optomechanically induced transparency (OMIT)~\cite{PhysRevA.81.041803,weis2010optomechanically,safavi2011electromagnetically,PhysRevA.90.043825,huang2014double,peng2020double}, directly analogous to electromagnetically induced transparency~\cite{PhysRevLett.62.1033,PhysRevLett.64.1107,marangos1998electromagnetically}. Asymmetric transmission profiles characteristic of Fano interference~\cite{PhysRevA.87.063813,PhysRevA.104.013708}. Their studies have linked these amplitude effects to a corresponding dispersive response: transparency regions are accompanied by positive group delay (slow light), while absorption features are associated with negative group delay (fast light). Introducing or tuning atomic rotation flips the local dispersion from slow to fast light, enabling controllable delay switching. Together, these established behaviors point toward applications in quantum state transfer~\cite{PhysRevLett.108.153604,PhysRevLett.108.153603}, orbital–angular–momentum metrology~\cite{zhang2021measuring}, low-power optical switching~\cite{stern2014fano,lu2017optomechanically}, and wavelength conversion~\cite{hill2012coherent}, which motivate the present study.

In Section II, we present the system and its Hamiltonian. Section III provides a brief discussion of the theoretical results. Finally, Section IV concludes the report with a summary of our work.
\section{MODEL AND HAMILTONIAN}\label{section:MODEL} 
\begin{figure}
        \centering
        \includegraphics[width=0.8\linewidth]{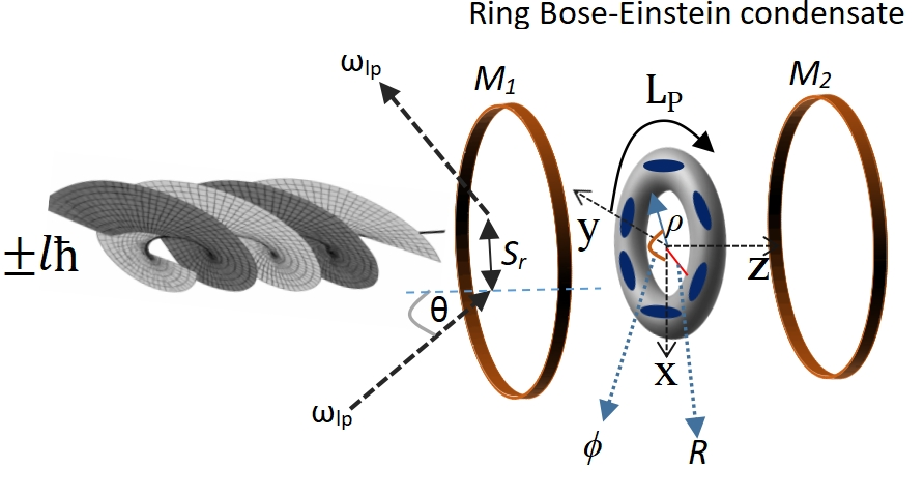}
        \caption{A Fabry-Pérot cavity with fixed mirrors $M_1$ and $M_2$, where the left mirror is driven by coherent control beams carrying OAM $\pm\hbar$, which interact with a toroidally trapped BEC. A second weak laser probes the cavity with an angle $\theta$ for the  investigation of the Goos-Hanchen Shift.}
        \label{figure1}
    \end{figure}

We consider our system configuration as depicted in Fig.~\ref{figure1}. At the core of this setup lies a BEC, composed of sodium (Na) of N atoms, confined within a toroidal trapping potential. This trap is symmetrically aligned within an optical cavity and can be realized with current experimental methods~\cite{PhysRevLett.99.083001,PhysRevA.83.043408,PhysRevLett.124.025301,PhysRevA.74.023617}.

The potential energy landscape experienced by an individual atom of mass $m$ in this ring-shaped geometry is given by

\begin{equation}
U(\rho, z) = \frac{1}{2}m\omega_\rho^2(\rho - R)^2 + \frac{1}{2}m\omega_z^2 z^2,
\end{equation}

where $R$ denotes the radius of the toroid, and $\omega_\rho$, $\omega_z$ represent the harmonic trapping frequencies in the radial and axial directions, respectively.

This potential effectively isolates the azimuthal degree of freedom $\phi$ from the radial ($\rho$) and axial ($z$) motions. We assume that the atomic wavefunction remains in its ground state in the $\rho$ and $z$ directions during the dynamics. As such, the relevant motion of the atoms is along the azimuthal direction, which is not subjected to any external confinement.

A one-dimensional treatment of the azimuthal dynamics is valid under current laboratory capabilities~\cite{
PhysRevA.74.023617}, and has proven to be accurate in describing experiments that incorporate three-dimensional aspects~\cite{PhysRevLett.110.025302,PhysRevLett.113.135302,PhysRevA.88.063633}. This approach is valid when the total number of atoms $n$ satisfies the condition~\cite{PhysRevLett.124.025301}

\begin{equation}
n < \frac{4\sqrt{\pi}R}{3 a_{\text{na}}} \left( \frac{\omega_\rho}{\omega_z} \right)^{1/2},
\end{equation}

where $a_{\text{na}}$ is the $s$-wave scattering length for sodium atoms.

To explore the dynamics of the BEC, we introduce a coherent light beam into the cavity. These beams are frequency-matched and share a common laser origin, but carry opposite orbital angular momenta $\pm l \hbar$. Their superposition forms a periodic intensity distribution around the cavity axis a so called angular optical lattice as demonstrated in prior studies~\cite{naidoo2012intra}. The light is blue-detuned from the atomic transition, minimizing absorption and spontaneous emission. Its effect is primarily through conservative dipole forces.

The rotational kinetic energy of the atoms is defined as
\begin{eqnarray}
 H_{\text{R,K}}&=&-\frac{\hbar}{2I_a}\int_0^{2\pi}\psi^\dagger(\phi)\frac{d^2}{d\phi^2}\psi(\phi)d\phi
\end{eqnarray}
here $I_a$ is the moment of inertia of the atoms about the cavity axis. We define the ansatz as
\begin{eqnarray}
    \psi(\phi)&=&\frac{\text{exp}[iL_p\phi]}{\sqrt2}c_p+\frac{\text{exp}[i(L_p+2l)\phi]}{\sqrt2}c_+\nonumber\\[8pt]&&+\frac{\text{exp}[i(L_p-2l)\phi]}{\sqrt2}c_-
\end{eqnarray}
\begin{eqnarray}
    \psi^\dagger(\phi)&=&\frac{\text{exp}[-iL_p\phi]}{\sqrt2}c_{p}^\dagger+\frac{\text{exp}[-i(L_p+2l)\phi]}{\sqrt2}c_+^\dagger\nonumber\\[8pt]&&+\frac{\text{exp}[-i(L_p-2l)\phi]}{\sqrt2}c_-^\dagger
\end{eqnarray}
using the ansatz from equation (4) and (5) in equation (3) we get 
\begin{eqnarray}
     H_{\text{R,K}}&=&\frac{\pi\hbar^2}{2I_a}(L_p^2c_p^\dagger c_{+}+(2l+L_p)^2c_{+}^\dagger c_{+}\nonumber\\[8pt]&&+(-2l+L_p)^2c_{-}^\dagger c_{-})
\end{eqnarray}
Introducing the operator $\sqrt{n}c=c_p^\dagger c_{+}$ and $\sqrt{n}d=c_p^\dagger c_{-}$ and apply to equation (6) we get
\begin{eqnarray}
     H_{\text{R,K}}&=&\frac{\hbar^2(L_p+2l)^2}{2I_a}c^\dagger c+\frac{\hbar^2(L_p-2l)^2}{2I_a}d^\dagger d
\end{eqnarray}
We define the particlelike excitations of the
condensate with their frequencies as follows.
\[
\omega_c = \frac{\hbar(L_p + 2l)^2}{2I_a}, \quad \omega_d = \frac{\hbar(L_p - 2l)^2}{2I_a}.
\]
and $H_{\text{R,K}}$ can be expressed as
\begin{equation}
    H_{\text{R,K}}=\hbar\omega_c c^\dagger c+\hbar\omega_d d^\dagger d
\end{equation}
Next we define the interaction of the atoms with the optical lattice
by the Hamiltonian

\begin{eqnarray}
    H_{A,L}&=&\int_0^{2\pi}\psi^\dagger(\phi)\hbar u_0\text{Cos}^2(l\phi)a^\dagger a\psi(\phi)d\phi
\end{eqnarray}
and applying the same ansatz given by equation (4) and (5) we get
\begin{eqnarray}
    H_{A,L}&=&\hbar u_0a^\dagger a(\frac{n}{2}+\frac{1}{4}(c_p^\dagger c_{+}+c_{+}^\dagger c_p\nonumber\\[8pt]&&+c_P^\dagger c_{-}+c_{-}^\dagger c_p))
\end{eqnarray}
and by using the operator $\sqrt{n}c=c_p^\dagger c_{+}$ and $\sqrt{n}d=c_p^\dagger c_{-}$ in equation (10) the Hamiltonian $H_{A,L}$ can be expressed as

\begin{eqnarray}
    H_{A,L}&=&\frac{\hbar u_0 n}{2}a^\dagger a+\frac{\hbar u_0\sqrt{n}\sqrt{2}}{4}a^\dagger a x_c\nonumber\\[8pt]&&+\frac{\hbar u_0\sqrt{n}\sqrt{2}}{4}a^\dagger a x_d
\end{eqnarray}

Next the atomic collisions are represented by the interaction term
\begin{equation}
H_{\text{Int}} = \frac{g}{2} \int_0^{2\pi} \psi^\dagger(\phi) \psi^\dagger(\phi) \psi(\phi) \psi(\phi) \, d\phi
\end{equation}
where \( g \) quantifies the atom-atom interaction strength.

Using the ansatz of Eq. (4) and (5), and the mode operators $\sqrt{n}c=c_p^\dagger c_{+}$ and $\sqrt{n}d=c_p^\dagger c_{-}$, this expression becomes
\begin{eqnarray}
    H_{\text{Int}}&=&\hbar g^\prime \big(n^2 - n + c^\dagger c c^\dagger c - c^\dagger c+ d^\dagger d d^\dagger d\nonumber\\[8pt]&&- d^\dagger d + 4n c^\dagger c+ 4n d^\dagger d + 2n c d^\dagger\nonumber\\[8pt]&&+ 2n c^\dagger d + 4 c^\dagger c d^\dagger d
\end{eqnarray}
Expressing Equation (13) in terms of the quadratures of modes $\sqrt{2}x_c=(c^\dagger+c)$ and $\sqrt{2}x_d=(d^\dagger+d)$
\begin{eqnarray}
H_{\text{int}}&=& \hbar g^\prime[N(X_c + iY_c)(X_d + iY_d) +\text{h.c.}\nonumber\\[8pt]&&+ 2N(X_c^2 + Y_c^2 - 1) + (X_d^2 + Y_d^2 - 1)\nonumber\\[8pt]&&+ (X_c^2 + Y_c^2 - 1)(X_d^2 + Y_d^2 - 1)\nonumber\\[8pt]&&+ \frac{1}{4} (X_c^2 + Y_c^2 - 1)(X_c^2 + Y_c^2 - 3)\nonumber\\[8pt]&&+ \frac{1}{4} (X_d^2 + Y_d^2 - 1)(X_d^2 + Y_d^2 - 3)
\end{eqnarray}
Where \( I_a \) is the moment of inertia of the system.
and
\begin{equation}
g^\prime = \frac{\omega_\rho a_s n_a}{2\pi R},
\end{equation}
where $\omega_\rho$ is the radial trapping frequency, $R$ is the radius of the trap and $a_{na}$ is the s-wave scattering length
between sodium atoms.

In the rotating frame of the driving laser, the effective Hamiltonian describing the azimuthal dynamics of the BEC takes the form of a one-dimensional interacting many-body system~\cite{PhysRevLett.111.043603,PhysRevLett.123.195301}.
\begin{eqnarray}
 H&=&-\frac{\hbar}{2I_a}\int_0^{2\pi}\psi^\dagger(\phi)\frac{d^2}{d\phi^2}\psi(\phi)d\phi\nonumber\\[8pt]&&+\int_0^{2\pi}\psi^\dagger(\phi)\hbar u_0\text{Cos}^2(l\phi)a^\dagger a\psi(\phi)d\phi\nonumber\\[8pt]&&+\frac{g}{2}\int_0^{2\pi}\psi^\dagger(\phi)\psi^\dagger(\phi)\psi(\phi)\psi(\phi)d\phi\nonumber\\[8pt]&&-\hbar\tilde\Delta a^\dagger a+i\hbar\eta_{lc}(a^\dagger-a)\nonumber\\[8pt]&&+i\hbar\eta_{lp}(a^\dagger e^{-i\delta t}-ae^{i\delta t})
\end{eqnarray}

So far, as we mentioned earlier that, the hybrid quantum system under investigation, illustrated schematically in Fig.\ref{figure1}, comprises The optical photons of the cavity that interact with a ring-shaped BEC that exhibits quantized circulation are confined to a toroidal trap inside the cavity~\cite{PhysRevLett.127.113601}. The cavity is driven by degenerate modes carrying OAM \(\pm l\hbar\), forming an angular lattice that couples to the azimuthal motion of the BEC. A second weak laser probe is incident with an angle $\theta$ in order to investigate the GHS in the transmitted or reflected field.

By applying the following quadrature to equation (8) and (11)
\begin{eqnarray}
   x_c&=&\frac{c^\dagger+c}{\sqrt{2}}, y_c=\frac{c-c^\dagger}{i\sqrt{2}}
\end{eqnarray}
\begin{eqnarray}
   x_d&=&\frac{d^\dagger+d}{\sqrt{2}}, y_d=\frac{d-d^\dagger}{i\sqrt{2}}
\end{eqnarray}
The total Hamiltonian, consisting of \( H_{R,K} \),\( H_{A,L} \), and the interaction term \( H_{\text{Int}} \), reads:

\begin{eqnarray}
 \mathcal{H}&=&\frac{\hbar\omega_c}{2}(x_c^{2}+y_c^{2})+\frac{\hbar\omega_d}{2}(x_d^{2}+y_d^{2})+\hbar Gx_c a^\dagger a+\hbar Gx_d a^\dagger a\nonumber\\[8pt]&&+2\hbar g^\prime n(x_c^{2}+y_c^{2})+2\hbar g^\prime n(x_d^{2}+y_d^{2})\nonumber\\[8pt]&&+2\hbar g^\prime n(x_cx_d-y_cy_d)-\hbar\tilde\Delta a^\dagger a\nonumber\\[8pt]&&+i\hbar \eta_{lc}(a^\dagger-a)+i\hbar\eta_{lp}(a^\dagger e^{-i\delta t}-ae^{i\delta t}).
\end{eqnarray}
\begin{figure*}
	\centering
	\includegraphics[width=0.8\linewidth]{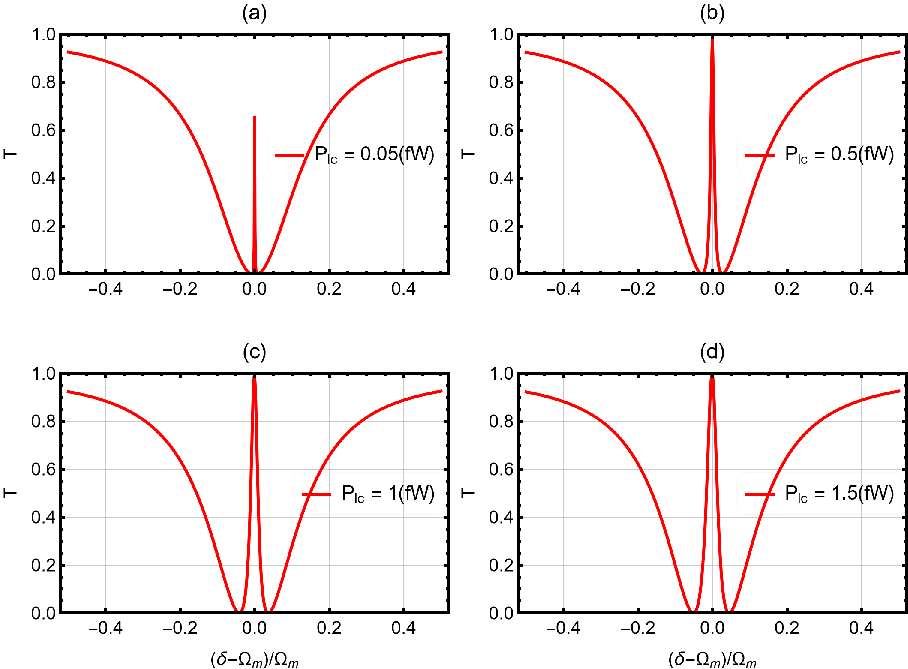}
	\caption{Transmission T of the probe field against the pump-probe detuning $(\delta-\Omega_m)/\Omega_m$, when $L_p=0$ for different Power.}
	\label{figure2}
\end{figure*}
\begin{figure*}
	\centering
	\includegraphics[width=0.8\linewidth]{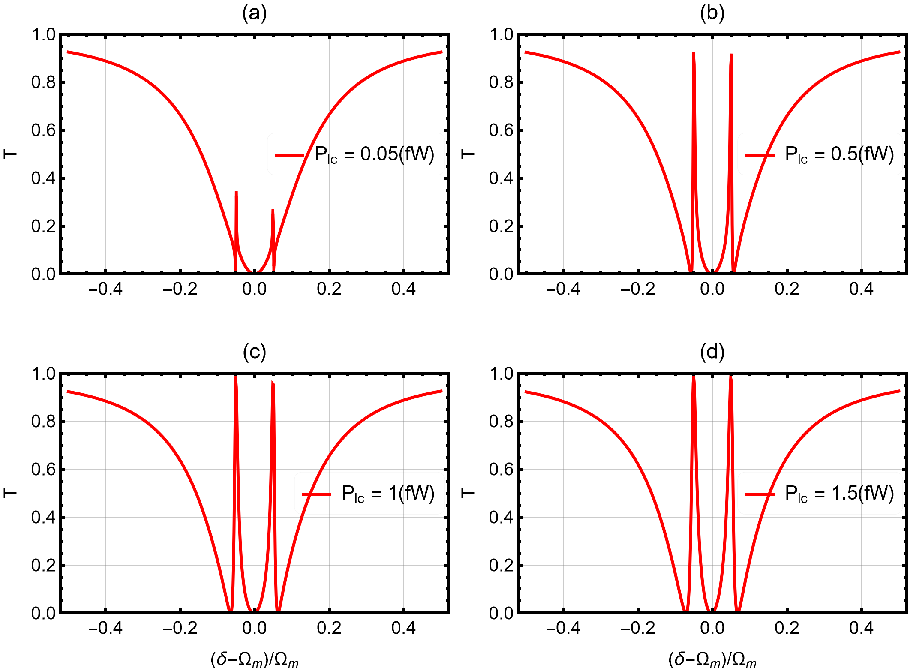}
	\caption{Transmission T of the probe field against the pump-probe detuning $(\delta-\Omega_m)/\Omega_m$, when $L_p=1$ for different Power.}
	\label{figure3}
\end{figure*}
\begin{figure*}
	\centering
	\includegraphics[width=0.8\linewidth]{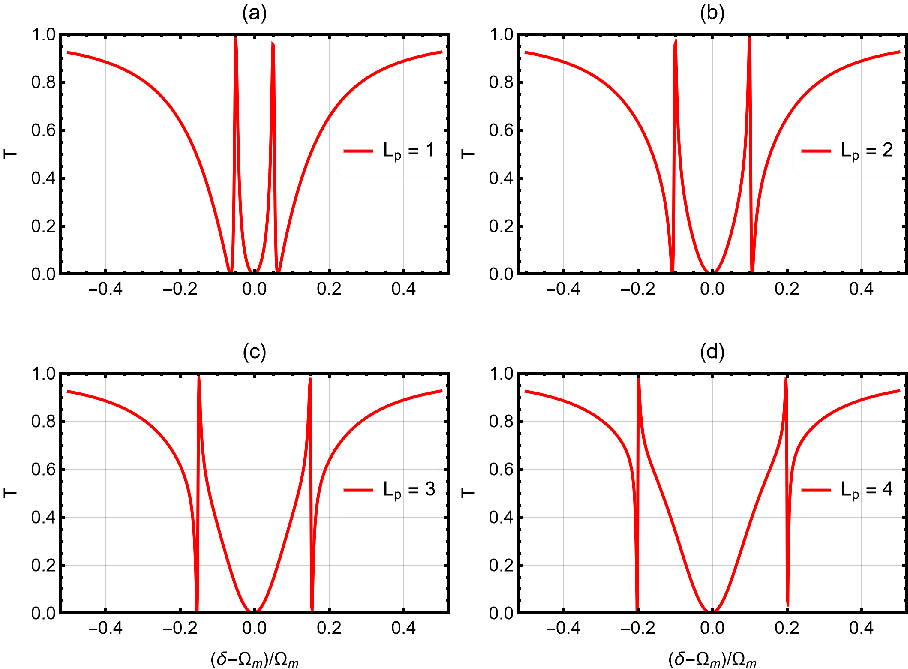}
	\caption{Transmission T of the probe field against the pump-probe detuning $(\delta-\Omega_m)/\Omega_m$, when $P{lc}=1$fW fixed, for different winding number.}
	\label{figure4}
\end{figure*}
\begin{figure*}
	\centering
	\includegraphics[width=0.8\linewidth]{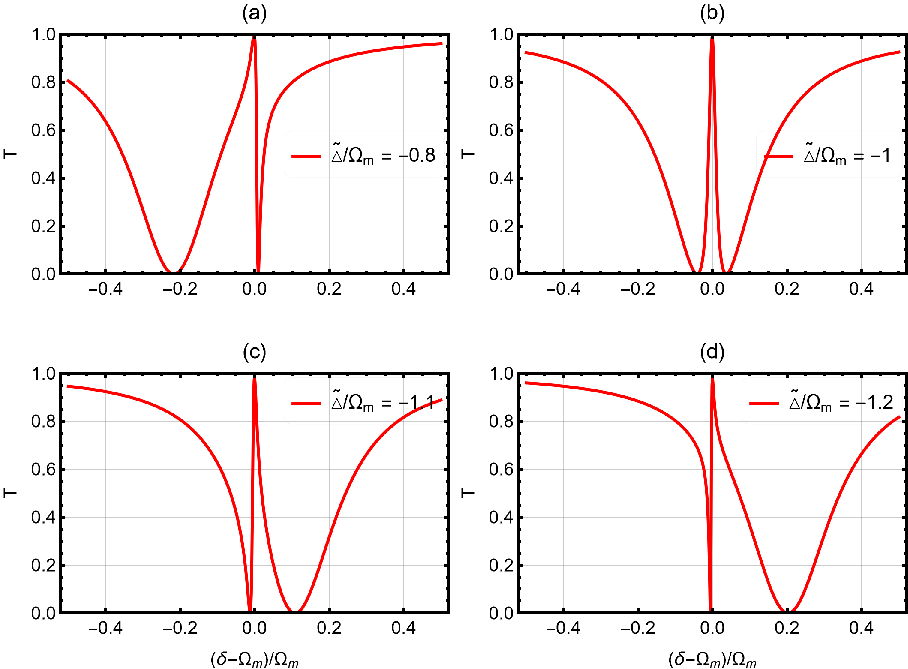}
	\caption{Transmission T of the probe field against the pump-probe detuning $(\delta-\Omega_m)/\Omega_m$, when $P{lc}=1$fW and $L_p=0$, for different cavity detuning $\tilde\Delta/\Omega_m$.}
	\label{figure5}
\end{figure*}
\begin{figure*}
	\centering
	\includegraphics[width=0.8\linewidth]{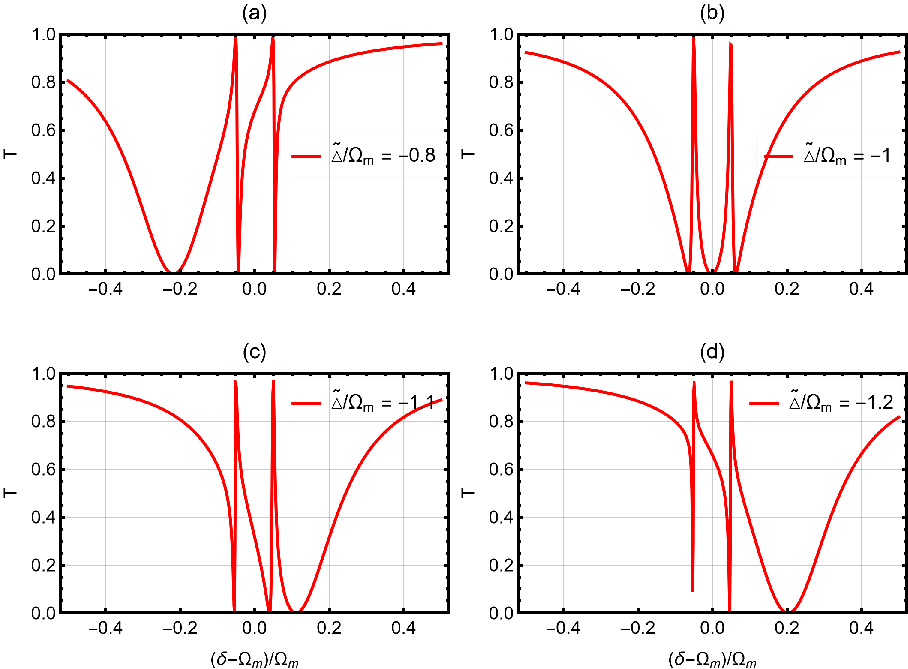}
	\caption{Transmission T of the probe field against the pump-probe detuning $(\delta-\Omega_m)/\Omega_m$, when $P{lc}=1$fW and $L_p=1$, for different cavity detuning $\tilde\Delta/\Omega_m$.}
	\label{figure6}
\end{figure*}
\begin{figure*}
	\centering
	\includegraphics[width=0.8\linewidth]{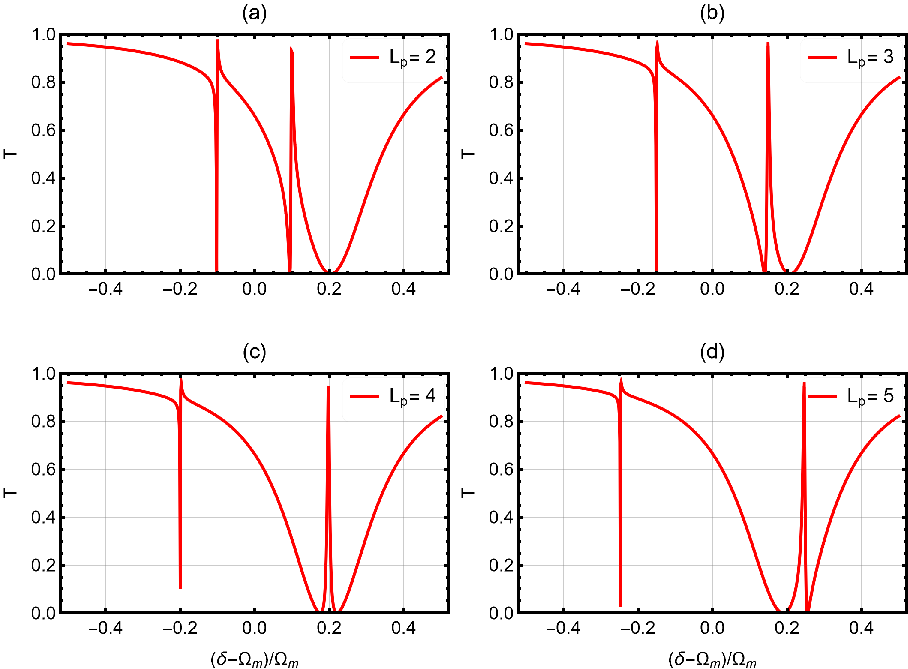}
	\caption{Transmission T of the probe field against the pump-probe detuning $(\delta-\Omega_m)/\Omega_m$, when $P{lc}=1$fW and $\tilde\Delta/\Omega_m=-1.2$, for different winding number $L_p$.}
	\label{figure7}
\end{figure*}
\begin{figure*}
	\centering
	\includegraphics[width=0.8\linewidth]{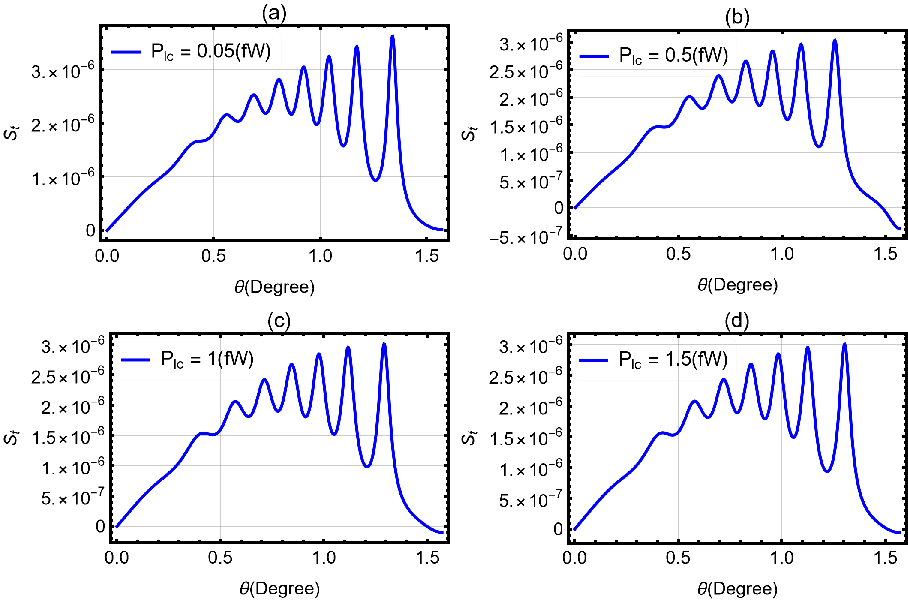}
	\caption{GHS $S_t$ of the transmission coefficient against incident angle $\theta$. when $L_p=0$, for different powers.}
	\label{figure8}
\end{figure*}
\begin{figure*}
	\centering
	\includegraphics[width=0.8\linewidth]{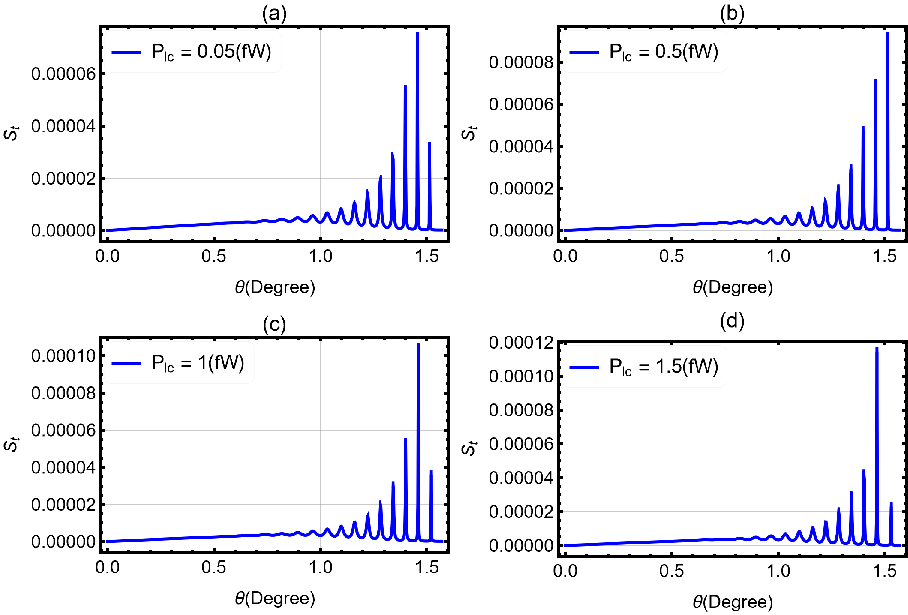}
	\caption{GHS $S_t$ of the transmission coefficient against incident angle $\theta$. when $L_p=1$, for different powers.}
	\label{figure9}
\end{figure*}
\begin{figure*}
	\centering
	\includegraphics[width=0.8\linewidth]{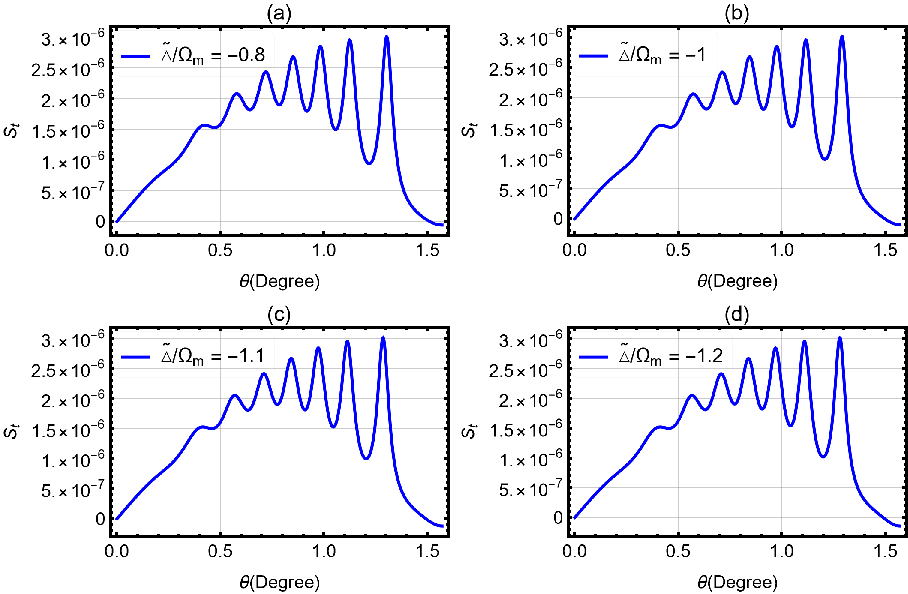}
	\caption{GHS $S_t$ of the transmission coefficient against incident angle $\theta$. when $L_p=0$, for different cavity detuning.}
	\label{figure10}
\end{figure*}
\begin{figure*}
	\centering
	\includegraphics[width=0.8\linewidth]{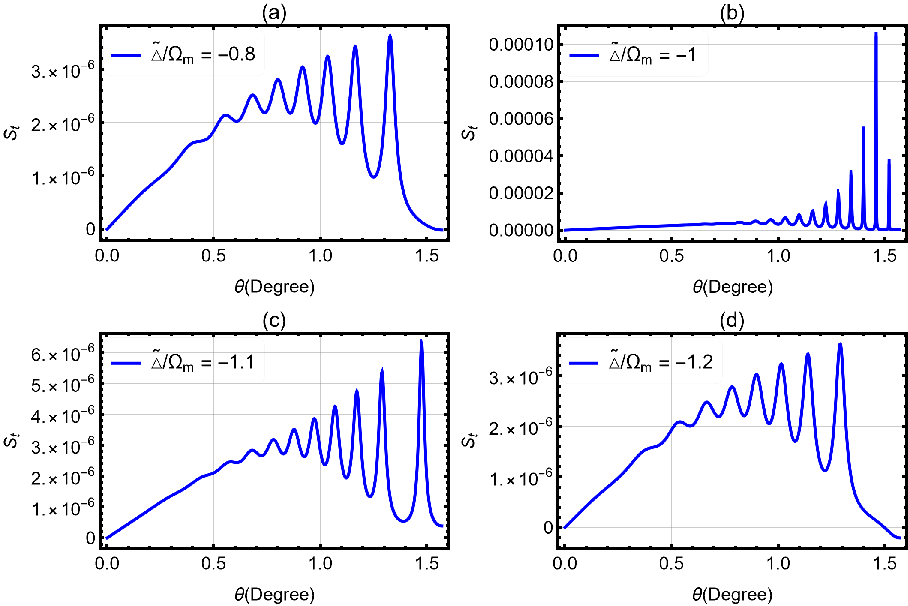}
	\caption{GHS $S_t$ of the transmission coefficient against incident angle $\theta$. when $L_p=1$, for different cavity detuning.}
	\label{figure11}
\end{figure*}
 In equation (19) $\eta_{lc}=\sqrt{\frac{\mu P_{lc}\gamma_0}{\hbar\omega_{lc}}}$ is the control laser field while $\eta_{lp}=\sqrt{\frac{\mu P_{lp}\gamma_0}{\hbar\omega_{lp}}}$ denotes the probe laser field, with the optical power $P_{lc}$, $P_{lp}$ and cavity line width $\gamma_0$. Where $\mu$ is the laser-cavity coupling parameter.Where $\delta=\omega_{lp}-\omega_{lc}$ is The detuning of the probe laser from the control laser.
From Bogoliubov theory, the corrected frequencies incorporating two-body atomic interactions are
\[
\omega_c' = \sqrt{\omega_c(\omega_c + 4 \tilde{g} N)}, \quad \omega_d' = \sqrt{\omega_d(\omega_d + 4 \tilde{g} N)},
\]
where \( {g^\prime} = \frac{g}{4\pi\hbar} \) and \( g = \frac{2\hbar\omega_\rho a_{\text{na}}}{R} \) denotes the effective 1D interaction strength, with \( a_{\text{na}} \) the atomic scattering length and \( \omega_\rho \) the radial trap frequency.

The  coupling between the side modes and the cavity field arises from the angular optical lattice and is quantified by
\[
G = \frac{u_0 \sqrt{n}}{2\sqrt{2}}, \quad u_0 = \frac{g_a^2}{\Delta_a},
\]
where \( g_a \) is the single-atom single-photon coupling strength and \( \Delta_a \) is the detuning between the optical field and atomic transition. The quadrature operators for the side modes are defined as~\cite{PhysRevLett.127.113601}

In addition to the BEC-cavity interaction, in equation (19), $\tilde\Delta = \omega_{lc} - \frac{g_{a}^2n}{2\Delta_a}$ represent the detuning of the cavity modes, respectively. The operators $a$, $a^\dagger$ are the annihilation and creation operators for the cavity modes.

 After considering the corresponding dissipation and fluctuating terms, the Heisenberg-Langevin equation~\cite{PhysRevLett.46.1} based on the total Hamiltonian in Eq. (19) can then be written as:
\begin{eqnarray}
\dot{a} &=& -\frac{\gamma_{0}}{2} -i(\tilde{\Delta} - G(x_c + x_d)) a+\eta_{lc} \nonumber\\[8pt]&&+ \eta_{lp} e^{-i \delta t} +\sqrt{\mu \gamma_0}\, a^{(\text{in})}
\end{eqnarray}
\begin{eqnarray}
\ddot{x}_c &=& -\gamma_m \dot{x}_c - \Omega_c^2 x_c - G \omega_{c}^\prime a^\dagger a - \mathcal{A} x_d + \omega_c \epsilon_c
\end{eqnarray}
\begin{eqnarray}
\ddot{x}_d &=& -\gamma_m \dot{x}_d - \Omega_d^2 x_d - G \omega_{d}^\prime a^\dagger a + \mathcal{A} x_c + \omega_d \epsilon_d.
\end{eqnarray}

Here, $\gamma_m$ denotes the mechanical damping rate of the side modes, while $a^{(\text{in})}$ accounts for the quantum fluctuations entering the cavity \cite{bowen2015quantum}. The stochastic terms $\epsilon_j$ represent the Brownian noise acting on the oscillators \cite{RevModPhys.86.1391}. For compactness, we have defined 
\begin{equation}
\Omega_c^2 = (\omega_c + 4g^2 n)^2 - 4g^2 n^2\nonumber
\end{equation}
\begin{equation}
\Omega_d^2 = (\omega_d + 4g^2 n)^2 - 4g^2 n^2\nonumber
\end{equation}
\begin{equation}
    \omega_{c}^\prime=\omega_c + 2gn, \omega_{d}^\prime=\omega_d + 2gn\nonumber
\end{equation}
\begin{equation}
    \mathcal{A} = 2g^\prime n(\omega_c - \omega_d)
\end{equation}

To analyze the Stokes field output, which is phase-matched with the probe frequency, we work in the limit where the probe drive is much weaker than the control drive ($\eta_c \gg \eta_p$). Under this condition, the weak probe can be treated as a perturbation to the otherwise steady-state dynamics of the cavity \cite{brennecke2008cavity}. Linearizing around this steady state allows us to obtain explicit expressions for the fluctuation amplitudes of the optical field.

The mean-field solutions can be written as
\begin{align}
\alpha &= \braket{a} = \alpha_s + A_{-} e^{-i\delta t} + A_{+} e^{i\delta t}, \\[6pt]
Q_c &= \braket{x_c} = Q_{cs} + C e^{-i\delta t} + C^\ast e^{i\delta t}, \\[6pt]
Q_d &= \braket{x_d} = Q_{ds} + D e^{-i\delta t} + D^\ast e^{i\delta t},
\end{align}
where $\alpha_s$, $Q_{cs}$, and $Q_{ds}$ are the steady-state components in the adiabatic limit ($\gamma_0 \gg \gamma_m, g_a^2/\Delta_a$). Explicitly, they are
\begin{align}
\alpha_s &= \frac{\eta_c}{\gamma_0/2 - i\Delta}, \label{eq:alphas}\\[6pt]
Q_{cs} &= -\frac{G\omega_{c}^\prime |\alpha_s|^2}{\sqrt{A^2 + \Omega_c^2 \Omega_d^2}}, \label{eq:qcs}\\[6pt]
Q_{ds} &= -\frac{G\omega_{d}^\prime |\alpha_s|^2}{\sqrt{A^2 + \Omega_c^2 \Omega_d^2}}, \label{eq:qds}
\end{align}
with effective detuning
\[
\Delta = \tilde{\Delta} - G(Q_{cs} + Q_{ds}).
\]

and

\begin{equation}
    \mathcal{G} = \frac{A \chi_c \chi_d (\omega_{c}^\prime - \omega_{d}^\prime) + \chi_c \omega_{c}^\prime + \chi_d \omega_{d}^\prime}{A^2 \chi_c \chi_d + 1}\nonumber
\end{equation}
Applying the input-output relation, the transmitted cavity field is
\begin{equation}
{a_{\text{out}}} = s_{lc} e^{-i \omega_{lc} t} + s_S e^{-i (\delta + \omega_{lc})t} + s_{AS} e^{-i (\delta - \omega_{lc})t}, \label{eq:aout}
\end{equation}
with
\begin{eqnarray}
    s_{lc} &=&s_{lc} = \sqrt{\mu \gamma_0} \alpha_s - \frac{\eta_c}{\sqrt{\mu \gamma_0}},\nonumber\\[8pt]&&s_S = \sqrt{\mu \gamma_0} A_{-} - \frac{1}{\sqrt{\mu \gamma_0}},s_{AS} = \sqrt{\mu \gamma_0} A_{+}
\end{eqnarray}

The transmission amplitude at the probe frequency is then
\begin{equation}
t_S = \frac{s_S}{ \sqrt{\mu \gamma_0}}, \label{eq:ts}
\end{equation}
and the transmission spectrum is given by
\begin{equation}
T = |t_S|^2 = \left| 1 - \mu \gamma_0 \right|^2. \label{eq:T}
\end{equation}
in this context we define the GHS for the transmitted probe beam as
\begin{equation}
    \text{S}_t=-\frac{\lambda}{2\pi\lvert\text{T}_t\rvert^2}[\text{Re T}_t\frac{\text{d}}{\text{d}\theta}\text{Im T}_t+\text{Im T}_t\frac{\text{d}}{\text{d}\theta}\text{Re T}_t],
\end{equation}
here the transmission coefficient $\text{T}_t$ can be derived by using the transfer matrix method~\cite{PhysRevA.77.023811,abbas2021enhancement} and can be written as
\begin{equation}
    \text{T}_t=\frac{2\text{Q}_0}{\text{Q}_0(\text{q}_{22}+\text{q}_{11})-(\text{Q}_{0}^2\text{q}_{12}+\text{q}_{21})}.
\end{equation}
In equation (36) $\text{Q}_0=\sqrt{\epsilon_0-\text{sin}^2\theta}$ and $\text{q}_{ij}$ are the elements of the transfer matrix $\text{q}(\text{k}_{z},\omega_\text{P})=\text{m}_1(\text{k}_{z},\omega_P,\text{d}_1)\text{m}_2(\text{k}_{z},\omega_P,\text{d}_2)\text{m}_1(\text{k}_{z},\omega_P,\text{d}_1)$ with (ij=1,2) where $\text{d}_1$ is the thickness of the mirror $\text{M}_1$, $\text{M}_2$ and $\text{d}_2$ is related to the intracavity with the Bose-Einstein condensate and OPA. The elements $\text{m}_j$ are related to the probe field and can be expressed as 
\begin{equation}
\text{m}_j(\text{k}_z,\omega_P,\text{d}_1)=\begin{bmatrix}
a_{11} & a_{12} \\
a_{21} & a_{22}
\end{bmatrix},
\end{equation}
with
\begin{eqnarray}
    &a_{11}&=\text{cos}(\text{k}^x_j\text{d}_j),\nonumber  \\[8pt]&& \nonumber  a_{12}=\text{i}\text{sin}(\text{k}^x_j\text{d}_j)\frac{\text{k}}{\text{k}^x_j},\\[8pt]&& \nonumber a_{21}=\text{cos}(\text{k}^x_j\text{d}_j), \\[8pt]&& \nonumber a_{22}=\text{i}\text{sin}(\text{k}^x_j\text{d}_j)\frac{\text{k}}{\text{k}^x_j}.\nonumber 
\end{eqnarray}
where $\text{k}^x_j=(\frac{\omega_\text{P}}{c})\sqrt{\epsilon_j-\text{sin}^2\theta}$.

\section{Results and discussion}
\label{Sec:results}
\subsection{Optomechanically Induced Transparency}
To investigate how the control field mediates interference in the probe output, we adopt experimentally attainable parameters~\cite{PhysRevLett.110.025302,PhysRevLett.127.113601,PhysRevA.74.023617,naidoo2012intra,de2021versatile}
atomic mass $m=23\,\mathrm{amu}$; ring radius $R=10\,\mu\mathrm{m}$; atom number $n=10^{4}$; and ${g}^\prime\approx 12\,{g}^\prime_{m}$.
We further set $\omega_{z}/2\pi=\omega_{\rho}/2\pi=840\,\mathrm{Hz}$.
With these values, the protocol is minimally invasive: Only a tiny fraction of atoms are diffracted by Bragg from the ring in first-order side modes.
As checked in Sec.~II, this parameter set (i) justifies a one–dimensional model, (ii) satisfies $\omega_{c,d}\gg 4{g}^\prime n$, and (iii) fulfills the weak–potential criterion of Ref.~\cite{PhysRevLett.127.113601}.
To demonstrate the effectiveness of rotation–controlled coherent interference, in what follows we numerically compute the probe–transmission spectrum; in those calculations we find that the interatomic coupling has a negligible impact on the transmission and therefore set ${g}^\prime=0$\cite{PhysRevLett.127.113601}.

\noindent When atomic rotation is absent ($L_p=0$), the two mechanical sidebands are degenerate.  
With the control beam applied, the probe transmission exhibits the hallmark of optomechanically induced transparency (OMIT) \cite{PhysRevA.81.041803}, as illustrated in Figure\ref{figure2}.  
The control and probe fields together drive the radiation–pressure force at the beat note $\delta=\omega_{lp}-\omega_{lc}$.  
If this drive is resonant with the mechanical frequency $\Omega_m$, the mechanical side mode oscillates coherently, producing Stokes $(\omega_{lc}-\Omega_m)$ and anti–Stokes $(\omega_{lc}+\Omega_m)$ scattering from the strong control field.  
At detuning $\tilde{\Delta}=-\Omega_m$, the Stokes process is far from resonance and is strongly suppressed.  
As a result, anti-Stokes photons, frequency-degenerate with the probe, accumulate in the cavity. Destructive interference between the probe field and this anti–Stokes scattering prevents a build-up of the intra-cavity probe field, opening a transparency window \cite{peng2020double,xiong2018fundamentals}.

\noindent The same physics can be interpreted using the $\Lambda$–type level diagram for $L_p=0$ Figure 3(b) of reference\cite{PhysRevA.107.013525}.  
where, the probe directly addresses the transition $\lvert 1\rangle \leftrightarrow \lvert 3\rangle$ while leaving the mechanical occupation unchanged. When the control laser is tuned near the red sideband, it enables an indirect pathway via the transition $\lvert 2\rangle \leftrightarrow \lvert 3\rangle$, in which a mechanical quantum is annihilated and a cavity photon is created.  
Because the control is much stronger than the probe, the excitation amplitude of this indirect route becomes comparable to that of the direct route.  
Under the resonant conditions $\delta=\Omega_m$ and $\tilde{\Delta}=-\Omega_m$, the indirect pathway acquires a total phase shift of $\pi$ (with $\lvert 2\rangle\!\to\!\lvert 3\rangle$ and $\lvert 3\rangle\!\to\!\lvert 2\rangle$ each contributing $\pi/2$), making it out of phase with the direct excitation.  
Consequently, the paths $\lvert 1\rangle\!\to\!\lvert 3\rangle$ and $\lvert 1\rangle\!\to\!\lvert 3\rangle\!\to\!\lvert 2\rangle\!\to\!\lvert 3\rangle$ destructively interfere, producing transparency for the probe.

\noindent These interference effects are typically analyzed in the linearized, essentially classical regime where quantum nonlinearities are neglected \cite{PhysRevLett.111.133601,PhysRevLett.111.053603}.  
In that limit, the OMIT response is governed by the combined parameter $G\lvert \alpha_s\rvert$.  
Classical OMIT is usually explored for $\gamma_m \ll \Gamma_{\mathrm{opt}}$ (with
$\Gamma_{\mathrm{opt}}=\tfrac{4 G^2 \lvert \alpha_s\rvert^2}{\gamma_o}$) and
$G\lvert \alpha_s\rvert \ll \gamma_o \ll \Omega_m$, so the width of the transparent window is set by $\Gamma_{\mathrm{opt}}$.

\noindent Because the control-laser power adjusts the degree of interference, the main OMIT features, such as the transmission peak and the window width, can be adjusted by varying the control field strength. In particular, the peak transmission saturates while the window broadens as the control power increases.

\noindent For a nonrotating condensate ($L_p=0$), the Bragg scattered mechanical side modes are degenerate and, in the presence of a control laser, the probe transmission displays OMIT.  
When a persistent current is present in the ring BEC ($L_p\neq0$), the side modes split and acquire distinct frequencies, which reshapes the probe transmission spectrum as shown in Figure \ref{figure3}.  
In particular, a double-OMIT profile appears: two transparency windows flank an absorption region whose width is approximately $\sim\!\lvert\Omega_c-\Omega_d\rvert$, centered at $\delta=\Omega_m$.  
The two transmission maxima occur at $\delta_{+}=\Omega_d$ and $\delta_{-}=\Omega_c$, respectively, where $\delta_{+}$ ($\delta_{-}$) denotes the peak on the positive (negative) side of $\Omega_m$.

\noindent The origin of this double-OMIT structure can be understood from interference among three excitation pathways that arise for $L_p\neq0$ in a double-$\Lambda$ configuration, shown in the right panel of Figure 3(b) of reference\cite{PhysRevA.107.013525}..  
As in the $L_p=0$ case, destructive interference produced by a $\pi$ phase difference between the pathways—(i) $\lvert1\rangle\!\to\!\lvert3\rangle$ together with $\lvert1\rangle\!\to\!\lvert3\rangle\!\to\!\lvert2_c\rangle\!\to\!\lvert3\rangle$, or (ii) $\lvert1\rangle\!\to\!\lvert3\rangle$ together with $\lvert1\rangle\!\to\!\lvert3\rangle\!\to\!\lvert2_d\rangle\!\to\!\lvert3\rangle$—suppresses the intracavity field at the corresponding beat notes.  
Hence, whenever the side mode frequencies associated with $\lvert2_c\rangle$ and $\lvert2_d\rangle$ match the beat frequency $\delta$, a transparency window opens, yielding the observed double-OMIT response.  
In contrast, complete absorption at $\delta=\Omega_m$ results from constructive interference between the direct path $\lvert1\rangle\!\to\!\lvert3\rangle$ and the two indirect routes $\lvert1\rangle\!\to\!\lvert3\rangle\!\to\!\lvert2_c\rangle\!\to\!\lvert3\rangle$ and $\lvert1\rangle\!\to\!\lvert3\rangle\!\to\!\lvert2_d\rangle\!\to\!\lvert3\rangle$, because the indirect paths contribute a total phase shift of $2\pi$ relative to the direct transition.

\noindent Persistent currents in the ring BEC further impact the probe spectrum, increasing the winding number enlarges the frequency separation between the Bragg scattered mechanical side modes, which in turn broadens the absorption window within the double-OMIT profile, as illustrated in Figure\ref{figure4}.

\noindent Furthermore, the characteristic features of OMIT originate from coherent interference between the anti–Stokes scattering channel and the probe field, which is most pronounced when the pump–probe detuning $\delta$ and the cavity detuning $\tilde{\Delta}$ are tuned to the mechanical resonance $\Omega_m$. When the cavity is detuned from the red–sideband condition ($\tilde{\Delta}\neq -\Omega_m$), interference between a resonant pathway and a nonresonant background gives rise to asymmetric Fano line shapes. For vanishing circulation ($L_p=0$), scanning $\tilde{\Delta}$ produces tunable Fano profiles centered at $\delta=\Omega_m$ as shown in Figure\ref{figure5}; the asymmetry diminishes upon approaching resonance and disappears at $\tilde{\Delta}=-\Omega_m$, where conventional OMIT is recovered. In the presence of quantized circulation ($L_p\neq 0$) in a ring BEC, the Bragg–scattered mechanical side modes become nondegenerate, and near $\tilde{\Delta}=-\Omega_m$ the probe transmission exhibits two resonances at $\delta\simeq\Omega_c$ and $\delta\simeq\Omega_d$, characteristic of a double–OMIT response see Figure\ref{figure3} and Figure\ref{figure4}. Away from resonance ($\tilde{\Delta}\neq -\Omega_m$) the Fano asymmetry reemerges, with the dominant contribution switching across the resonance: for $-\tilde{\Delta}<\Omega_m$ the response is governed mainly by the $\Omega_c$ branch, whereas for $-\tilde{\Delta}>\Omega_m$ it is dominated by the $\Omega_d$ branch; the asymmetry vanishes at $\tilde{\Delta}=-\Omega_m$ as shown in Figure\ref{figure6}. Moreover, increasing the winding number $L_p$ enhances the splitting $|\Omega_c-\Omega_d|$, thereby enlarging the frequency separation between the corresponding absorption dips in the double–OMIT spectrum can be seen in Figure\ref{figure7}.
\subsection{Goos-H\"{a}nchen Shift}
We employ additional parameters $\epsilon_0=1$, $\epsilon_1=2.2$, mirror thickness $\text{d}_1=0.2\times10^{-6}\text{m}$, $\text{d}_2=5\times10^{-6}\text{m}$ \cite{PhysRevA.77.023811} for the investigation of the GHS. 
As the atomic rotation $L_p=0$, the Goos–Hänchen shift (GHS) of the transmitted probe, evaluated from Equation.~(35), behaves as follows across Figure\ref{figure8}(a-d). At $P_{lc}=0.05~\mathrm{fW}$ Figure\ref{figure8}(a) the cooperativity is small\cite{PhysRevA.107.013525} and the OMIT window is narrow and essentially symmetric, so the phase slope is steep and positive and the GHS increases monotonically from $0$ to $\sim 3\times10^{-6}$ without any negative part. At $P_{lc}=0.5~\mathrm{fW}$ Figure\ref{figure8}(b) the window is wider producing a short interval with negative angular phase slope and therefore a shallow dip to $\approx -5\times10^{-7}$ before the GHS increases to the same positive maximum $\approx 3\times10^{-6}$. At $P_{lc}=1~\mathrm{fW}$ Figure\ref{figure8}(c) and $P_{lc}=1.5~\mathrm{fW}$ Figure\ref{figure8}(d) the cooperativity is larger and the transparency is broader (flatter dispersion), but the drive is effectively on the red sideband at $L_p=0$, so the lineshape is symmetric and the GHS remains positive throughout, again climbing from $0$ to $\sim 3\times10^{-6}$. The similar peak value in all panels reflects that each curve samples the steepest available phase region and that the reduction of phase slope with power is partly compensated by the $|T_t|^{-2}$ factor in Eq.~(35), whereas the transient sign change at $P_{lc}=0.5~\mathrm{fW}$ originates from the small detuning offset rather than from atomic rotation (which is fixed at $L_p=0$).

Furthermore, with $L_p\neq 0$, we evaluate the Goos–Hänchen shift, $S_t$ and observe a monotonic growth of the maximum GHS with pump power, from $0\!\to\!6\times10^{-5}$ at $P_{lc}=0.05~\mathrm{fW}$ Figure\ref{figure9}(a), to $0\!\to\!8\times10^{-5}$ Figure\ref{figure9}(b) at $P_{lc}=0.5~\mathrm{fW}$, to $0\!\to\!1.0\times10^{-4}$ Figure\ref{figure9}(c) at $P_{lc}=1~\mathrm{fW}$, and to $0\!\to\!1.2\times10^{-4}$ Figure\ref{figure9}(d) at $P_{lc}=1.5~\mathrm{fW}$. Physically, the nonzero rotation biases the probe’s dispersive phase response (fixing the sign of the dominant phase slope) and enhances the phase contrast created by optomechanical interference; as the pump increases, the linearized coupling  boosts the cooperativity, which strengthens the rotation–biased dispersion and increases $T_t$ over the frequency region sampled by the angular scan. Consequently $S_t$ grows with $P_{lc}$ in this regime, and—unlike the $L_p=0$ case. Within the power range shown, power broadening does not yet dominate the phase-slope enhancement, so the peak GHS increases approximately with $C(P_{lc})$; at still higher powers one expects eventual saturation when broadening flattens the dispersion.

Moreover, for $L_p=0$ and varying cavity detuning $\tilde{\Delta}$. Keeping the control power fixed at $P_{lc}=1~\mathrm{fW}$ as in Figure\ref{figure5}, we vary the normalized cavity detuning $\tilde{\Delta}/\Omega_m\in\{-0.8,-1.0,-1.1,-1.2\}$ Figure\ref{figure10}(a-d) and compute the Goos–Hänchen shift, $S_t$. Figure\ref{figure5} shows that for $L_p=0$ the probe transmission exhibits tunable Fano asymmetry when the cavity is off the red sideband, while the asymmetry reduces as resonance is approached and disappears at $\tilde{\Delta}=-\Omega_m$, where standard OMIT occurs; these data are taken at $P_{lc}=1~\mathrm{fW}$, matching our conditions. In our GHS-vs-$\theta$ plots this manifests mainly as a horizontal shift of the angle at which the phase slope is steepest, not as a change in its peak magnitude: since $S_t$ is set by geometry, the maximum attainable $|T_t|$ at fixed power is essentially controlled by the linewidth/contrast of the transparency/Fano feature, which (for $L_p=0$) is set by cooperativity and thus by power rather than by small detuning offsets. Consequently, for all four detunings the GHS traces start at $0$ and rise to the same positive cap $\approx 3\times 10^{-6}$, while off-resonant cases ($\tilde{\Delta}/\Omega_m=-0.8,-1.1,-1.2$) chiefly translate the extremum in $\theta$ without appreciably altering its height. This interpretation is consistent with our paper’s account that tuning $\tilde{\Delta}$ at $L_p=0$ generates asymmetric Fano line shapes away from resonance but restores symmetric OMIT at $\tilde{\Delta}=-\Omega_m$ (hence comparable peak phase slopes at fixed power), and with the discussion that the phase dispersion governing group delay—and therefore our angular phase derivative—is the key driver of the effect. 

Next with atomic rotation turned on ($L_p\neq 0$) and the control power fixed at $P_{lc}=1~\mathrm{fW}$, the Goos–Hänchen shift $S_t$ is computed, and its detuning dependence follows the rotation–modified dispersion illustrated in Figure\ref{figure6}. Rotation splits the side modes and creates a central absorption, yielding two steep dispersive flanks. Exactly at the red sideband, $\tilde{\Delta}/\Omega_m=-1$, these flanks are symmetric and maximally steep, so the GHS rises from $0$ to a much larger value, $\approx 1.0\times 10^{-4}$. When the cavity is detuned to $\tilde{\Delta}/\Omega_m=-1.1$, the angular scan overlaps only one (partially shifted) steep flank, so the phase slope is smaller and the GHS reaches an intermediate maximum $\approx 6\times 10^{-6}$. Farther from resonance, at $\tilde{\Delta}/\Omega_m=-1.2$ and $-0.8$, the doublet is displaced such that the steepest phase regions are sampled even less effectively and the dispersion itself is weaker, so the GHS increases only to $\approx 3\times 10^{-6}$ in each case. In short, for $L_p\neq 0$ and fixed power, the peak GHS is set by how closely the angular scan samples the rotation–split transparency flanks: it is maximal at $\tilde{\Delta}=-\Omega_m$ and diminishes as $\tilde{\Delta}$ moves away, producing the sequence $3\times10^{-6}\,(-0.8)\ \rightarrow\ 1.0\times10^{-4}\,(-1)\ \rightarrow\ 6\times10^{-6}\,(-1.1)\ \rightarrow\ 3\times10^{-6}\,(-1.2)$ shown in Figure\ref{figure11}.

\section{Conclusion}
We have established a rotation–tunable Goos--H\"{a}nchen shift of the transmitted probe in a ring Bose--Einstein condensate coupled to an optical cavity, highlighting how optomechanical interference enables large, controllable beam displacements at ultralow powers. Without circulation, conventional optomechanically induced transparency provides a strictly positive and bounded shift whose magnitude is set primarily by cooperativity; small detuning offsets can introduce a brief sign reversal consistent with a weak Fano asymmetry. With circulation, the Bragg side modes split and the transmission develops paired transparency windows separated by a central absorption. The associated steep dispersive flanks bias and amplify the phase derivative, yielding a pronounced, power–tunable shift. 

At fixed power, the shift is maximized at the red–sideband condition and diminishes away from resonance, reflecting how effectively the angular scan samples the rotation–split dispersion. Increasing the winding number broadens the central absorption and steepens accessible phase gradients, further enhancing the attainable displacement. The protocol remains minimally invasive under realistic conditions, and interatomic interactions have negligible influence on the transmission features that set the phase slope.

Together, circulation, control power, and cavity detuning emerge as practical, in situ knobs for engineering the Goos--H\"{a}nchen shift, enabling interferometric beam steering and phase–gradient metrology in hybrid atom–optomechanical platforms. Future work can optimize mirror–stack geometry and operating points, extend beyond mean–field to quantify quantum and thermal noise, explore nonclassical probes, and pursue integrated implementations for compact rotation sensing.

\section*{Acknowledgements}
This work was supported by the National Natural Science Foundation of China (Grant No. 12174301), the Natural Science Basic Research Program of Shaanxi (Program No. 2023-JC-JQ-01), and the Fundamental Research Funds for the Central Universities.

%\bibliographystyle{apsrev4-2}
%\bibliography{biblio}% Produces the bibliography via BibTeX.
%
\bibliographystyle{apsrev4-2}
\bibliography{biblio}

\end{document}